\documentstyle[psfig]{caosp}
\begin{document}

\title{Mercury and platinum abundances in mercury-manganese stars}

\author{C. M. Jomaron\inst{1,}\and M. M. Dworetsky\inst{1} \and D. A. Bohlender \inst{2}}

\institute{University College London, Gower Street, London, England WC1E 6BT 
\and National Research Council of Canada, Herzberg Institute of
Astrophysics, 5071 W. Saanich Road, Victoria, BC, Canada V8X 4M6 }

\maketitle

\def\hgi{Hg\,{\sc i}\,\,} 
\def\hgii{Hg\,{\sc ii}\,\,} 
\def\hgiii{Hg\,{\sc iii}\,\,} 
\def\ptii{Pt\,{\sc ii}\,\,} 
\newcommand{\vsini}{$v\sin\!i$}

\begin{abstract}

We report new results for the elemental and isotopic abundances of the
normally rare elements mercury and platinum in HgMn stars. Typical
overabundances can be 4 dex or more.  The isotopic patterns do not follow
the fractionation model of White et al (1976).

\keywords{stars: abundances --- stars: chemically peculiar}

\end{abstract}

\section{HgMn stars}

The HgMn stars correspond in $T_{\rm eff}$ to the main-sequence
between A0 and B6 (11\,000--16\,000\,K). Their abundance anomalies
include both overabundances (e.g., Mn, P, Ga, Sr, Hg, Pt) and
underabundances (e.g., He, Al, Ni, Co). They have no detectable
ordered magnetic fields like the classical Ap stars (SrEuCr or Si
types). While normal A0 stars have a typical rms \vsini\,$\sim$\,164\,km\,s$^{-1}$ (Dworetsky, 1974), HgMn stars have typical \vsini\ of
10--20\,km\,s$^{-1}$; some are as low as 2--3\,km\,s$^{-1}$. Smith (1996)
is recommended for a more detailed overview.

\section{Data}

We obtained high-resolution, high $S/N$ optical spectra of \hgii and
\ptii lines for several stars with the Gecko spectrograph (R$\sim$10$^5$)
on the Canada-France-Hawaii Telescope, and complete spectra for many more
stars with the Hamilton \'{E}chelle Spectrograph
($\lambda\lambda$3900--9000, R$\sim$ 5$\times$10$^4$) at Lick Observatory. 

The models for the lines are based on isotopic and hyperfine structure
measurements from the literature (Engleman 1989; Dworetsky et al. 1998)
and on {\it gf-}values from Dworetsky et al (1984) and Dworetsky
(1980). We used the spectrum-synthesis codes UCLSYN (Smith 1992) and
BINSYN (Smalley 1996). Best fits for two stars, HR 7775 and $\chi$
Lupi, are shown in Fig.~\ref{fig1}, and the results for those and
several other stars are summarised in Tables~\ref{tab1}--\ref{tab3},
where abundances are given on the scale with $\log N({\rm H})=12$.

\section{\hgi vs \hgii}

The strongest optical lines of \hgi in these stars derive from the
$^3$S to $^3$P transitions at $\lambda$4046,
$\lambda$4358 and $\lambda$5461. Although the first of these is a
contaminant of the \ptii $\lambda$4046 feature, the other two are
clearly present in several stars.

Table~\ref{tab1} shows the abundance of \hgi derived for five stars using
the lines $\lambda$4358 and $\lambda$5461 and compares them with the
abundance derived for \hgii using the summed isotopes of the $\lambda$3984
line (CFHT data).  The agreement is excellent. 

\begin{table}\small
\begin{center}
\caption{Equivalent widths and abundances of Hg in five HgMn stars}
\label{tab1}
\begin{tabular}{c|cc|cc|cc}
Star & \multicolumn{2}{c}{\hgi $\lambda$4358} & \multicolumn{2}{c}{\hgi $\lambda$5461}& \multicolumn{2}{c}{\hgii $\lambda$3984}\\
&           log A & W(m\AA) &    log A & W(m\AA)&    log A& W(m\AA)\\
\hline
HR 7775    &     6.28 &   13     &  6.45  &  14   &    6.30 & 71\\
28 Her     &     5.60 &   3.5    &  5.80  &  4.5  &    5.72 & 28\\
$\phi$ Her  &     5.90 &   4      &  6.10  &  6	  &    6.20 & 66\\
$\iota$ CrB &     6.05 &   5      &  6.28  &  7.5  &    6.10 & 70\\
$\chi$ Lup  &     6.01 &   7.5    &  $<$6.50  &  12.5bl&    6.33 & 46\\
\hline
\end{tabular}\end{center}\end{table}

\section{Fractionation of Hg and Pt}

The stellar abundances of Hg and Pt are enormous compared to the
terrestrial (cosmic) abundances of Hg and Pt of 1.1 dex and 1.8 dex
respectively. Even more remarkably, as noted by White et al (1976),
Cowley \& Aikman (1975), and extended by Smith (1997) the abundance
pattern for Hg is skewed to the heavier isotopes in the cooler HgMn
stars. We confirm their results and extend them to Pt, which shows a
similar behaviour.

\begin{figure}
\begin{center} 
\vspace*{-10mm}
\leavevmode 
\psfig{figure=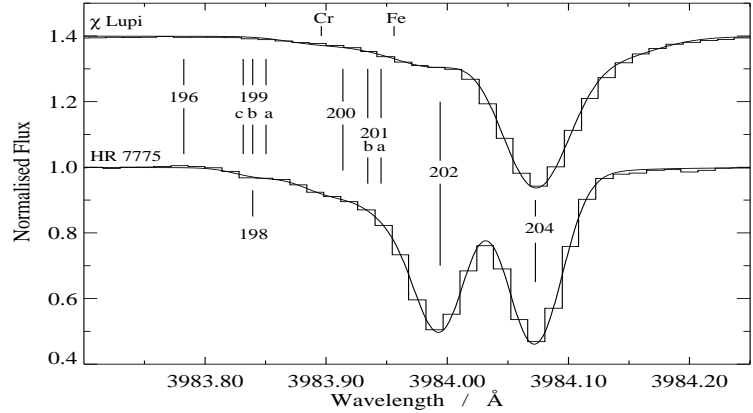,width=110mm,height=60mm}
\caption{Comparison of \hgii profiles in HR 7775 and $\chi$ Lupi: $\lambda$3984}
\label{fig1}
\end{center} 
\end{figure}

\begin{figure} 
\begin{center} 
\vspace*{-10mm}
\leavevmode 
\psfig{figure=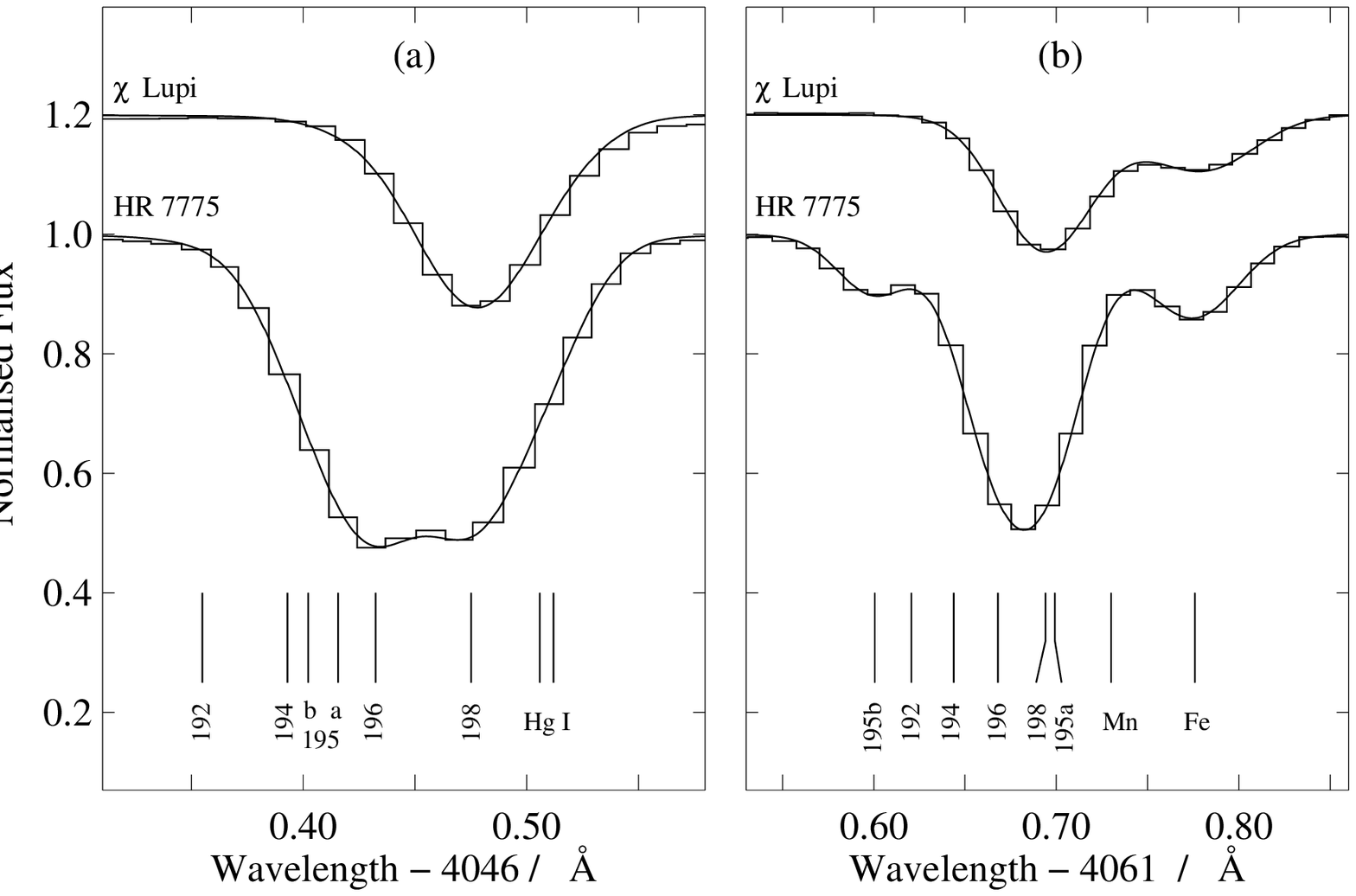,width=110mm,height=65mm}
\caption{Comparison of \ptii profiles in HR 7775 and $\chi$
Lupi: (a)$\lambda$4046; (b)$\lambda$4061 }
\label{fig2}
\end{center} 
\end{figure}

Tables~\ref{tab2}--\ref{tab3} show the isotope percentages of Hg and
Pt in several stars. Whilst the Hg in $\chi$ Lupi and 28 Her is
concentrated over 99\% into $^{204}$Hg, that in HR 7775, $\phi$ Her and
$\iota$ CrB is not nearly as concentrated into the heavy isotopes
(albeit much greater than the terrestrial mixture). The Pt anomalies
can be seen to `shadow' the Hg anomalies; stars with strong $^{204}$Hg
also have strong $^{198}$Pt. Figs.~\ref{fig1} \& \ref{fig2}
illustrate the differences between HR 7775 and $\chi$ Lupi for the
\hgii and \ptii lines.

\begin{table}\small
\begin{center}
\caption{Percentage Hg composition by isotope}
\label{tab2}
\begin{tabular}{c|cccccc}
isotope& $\iota$ CrB&HR 7775&$\phi$ Her&$\chi$ Lupi&28 Her&terrestrial\\
\hline
204       & 46.4 &  61.7   &89.1   & 98.8     	& $>$ 99& 6.87   \\
202       & 38.6 &  37.2   &10.1   & 1.1      	&$<$1	& 29.86  \\
201       &  7.7 &  0.4    & ?     & 0.1      	&-	& 13.18  \\
200       &  3.7 &  0.3    & ?     & -  	&-	& 23.10  \\
199       &  1.8 &  0.2    & ?     & -  	&-	& 16.87  \\
198       &  1.6 &  0.2    & ?     & -  	&-	& 9.97   \\
196       &  0.1 &   -     & -     & -  	&-	& 0.15   \\
\hline
\end{tabular}\end{center}\end{table}

\begin{table}\small
\begin{center}
\caption{Percentage Pt composition by isotope}
\label{tab3}
\begin{tabular}{c|ccc}
isotope& HR 7775&$\chi$ Lupi&terrestrial\\
\hline
198   & 45	&   92   & 7.2  \\
196   & 43	&    3 	 & 25.3  \\
195   & 11	&    ?  & 33.8  \\
194   & 1	&    -	 & 32.9  \\
192   & -	&    -	 & 0.8  \\
\hline
\end{tabular}\end{center}\end{table}

Leckrone et al (1993) found that $\chi$ Lupi \hgiii lines yielded a
similar isotopic anomaly to the \hgii$\lambda$3984 line, which excludes
the possibility that the light isotopes were `hiding' in a highly
fractionated cloud of \hgiii  (Michaud, Reeves \& Charland, 1974).

White et al proposed an extension of the radiative diffusion
vs. gravitational settling model, {\it fractionation}, to explain this
result. If one starts with the terrestrial mixture, by analogy with
mineralogy, it can be fractionated by a constant factor $\exp (q)$ (per
atomic mass unit). A strongly positive value of $q$ (e.g., 3) would push
99\% of Hg into $^{204}$Hg, normally a rare isotope. The parameter $q$
is derived from the equation

\begin{equation}
q = \frac{log\,\alpha}{log\,e\,(A-202)}
\end{equation}

where

\begin{equation} 
\alpha = \frac{ [N_{A}/N_{202}]_\star }{ [N_{A}/N_{202}]_\odot }
\end{equation}

A similar equation based on $^{196}$Pt may be defined for Pt isotopes.

\section{Results}

\begin{figure} 
\begin{center} 
\vspace*{-2mm}
\leavevmode 
\psfig{figure=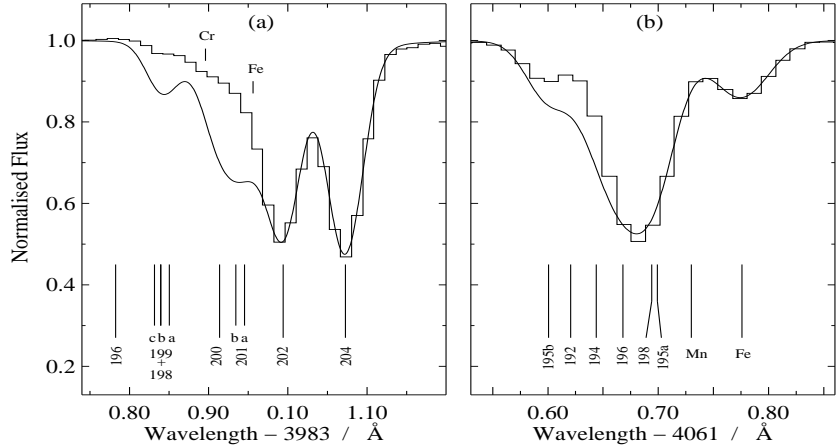,width=110mm,height=65mm}
\caption{Observed data versus synthetic fractionation profile for
HR 7775: (a) \hgii $\lambda$3984; (b) \ptii $\lambda$4061 }
\label{fig3}
\vspace*{-3mm}
\end{center} 
\end{figure}

The fractionation hypothesis does not adequately represent the best
observed structures for Hg and Pt. Fig.~\ref{fig3} shows the predicted
profiles (in HR 7775) for \hgii $\lambda$3984 and \ptii $\lambda$4061
for a constant $q$ as deduced from the abundance ratios of the two
heaviest isotopes. There is clearly not enough of the light isotopes
present.  In the cases of $\chi$ Lupi and 28 Her we find that the
enhancements of the heavy isotope of Hg are so extreme that virtually
all the Hg is in one isotope.

\section{Discussion}

Our results contradict what was previously a largely accepted concept,
namely, that the strong Hg and Pt isotopic abundance anomalies could be
parameterised in a fractionation model. This is clearly not the case in at
least several stars.

One (unlikely) {\it ad hoc} explanation is that the original
abundances in these stars were different from cosmic. The fact that
the Pt anomalies `shadow' the Hg may support this view. However, while
surely involving radiative diffusion and gravitational settling, any
complete theory needs to take into account other factors such as the
detailed flux profile of the stellar atmosphere. These exotic
abundances are now established results in need of a theoretical
explanation.

\acknowledgements
The authors gratefully acknowledge discussions with Rolf Engleman, Jr,
and contributions by students N.R. Crawley and D.J. Harman at UCL.
Research on chemically peculiar stars at UCL is supported by PPARC
grant GR/K58500 and travel to telescopes by PPARC grant GR/K60107.

\end{document}